\documentclass[12pt]{article}
\usepackage{amsmath}
\usepackage{graphicx}
\usepackage{eufrak}
\newlength{\dummysp}
\newcommand{\tr}{\mathop{{\hbox{Tr} \, }}\nolimits}

\newcommand{\stxt}[1]{\mathop{\hbox{{\scriptsize #1}}}\nolimits}
\newcommand{\bbar}[1]{{\overline{#1}}}
\newcommand{\half}{{\frac{1}{2}}}

\newcommand{\beq}{\begin{eqnarray}}
\newcommand{\eeq}{\end{eqnarray}}
\newcommand{\nnn}{ \nonumber \\ }

\newcommand{\e}{{\epsilon}}
\newcommand{\s}{{\sigma}}

\newcommand{\vev}[1]{{\langle #1 \rangle}}
\newcommand{\bigvev}[1]{{\left\langle #1 \right\rangle}}

\newcommand{\gappeq}{\mathrel{\rlap {\raise.5ex\hbox{$>$}}
{\lower.5ex\hbox{$\sim$}}}}
\newcommand{\lappeq}{\mathrel{\rlap{\raise.5ex\hbox{$<$}}
{\lower.5ex\hbox{$\sim$}}}}
\newcommand{\myref}[1]{(\ref{#1})}

\newcommand{\ben}{\begin{enumerate}}
\newcommand{\een}{\end{enumerate}}

\newcommand{\sqtw}{\sqrt{2}}

\newcommand{\bit}{\begin{itemize}}
\newcommand{\eit}{\end{itemize}}
\newcommand{\obf}{{\bf 1}}
\newcommand{\nbf}{{\bf n}}
\newcommand{\mbf}{{\bf m}}
\newcommand{\xb}{{\bbar{x}}}
\newcommand{\yb}{{\bbar{y}}}
\newcommand{\zb}{{\bbar{z}}}
\newcommand{\tmr}{t^{\mu \nu \rho}}
\newcommand{\tmn}{t^{\mu \rho \nu}}
\newcommand{\ibf}{\boldsymbol{\hat \imath}}
\newcommand{\jbf}{\boldsymbol{\hat \jmath}}

\newcommand{\Ncal}{{\cal N}}

\def\[{\left [}
\def\]{\right ]}
\def\({\left (}
\def\){\right )}

\begin{document}

\begin{titlepage}

\renewcommand{\thefootnote}{\fnsymbol{footnote}}

\hfill Sept.~15, 2003

\hfill hep-lat/0307024

\vspace{0.25in}

\begin{center}
{\bf \Large The fermion determinant \\
\vskip 10pt
in (4,4) 2d lattice super-Yang-Mills}
\end{center}

\vspace{0.15in}

\begin{center}
{\bf \large Joel Giedt\footnote{{\tt giedt@physics.utoronto.ca}}}
\end{center}

\vspace{0.15in}

\begin{center}
{\it University of Toronto \\
60 St. George St., Toronto ON M5S 1A7 Canada}
\end{center}

\vspace{0.15in}

\begin{abstract}
We find that the fermion determinant is not generally
positive in a class of lattice actions
recently constructed by Cohen et al.~[hep-lat/0307012];
these are actions that contain an exact lattice supersymmetry
and have as their target (continuum) theory
(4,4) 2-dimensional super-Yang-Mills.
We discuss the implications of this
finding for lattice simulations and give some
preliminary results for the phase of the determinant
in the phase-quenched ensemble.

\end{abstract}

\end{titlepage}

\renewcommand{\thefootnote}{\arabic{footnote}}
\setcounter{footnote}{0}

\section{Introduction}
A major motivation for efforts to latticize supersymmetric
models is that some nonperturbative aspects of supersymmetric field
theories are not accessible by the usual techniques,
most often relying on holomorphy.  Indeed, phenomenological applications
of softly broken $\Ncal=1$ 4d supersymmetric gauge theories such
as the {\it minimal supersymmetric standard model}
are generally regarded as effective field theories valid
for the TeV regime, derived from a more fundamental
supersymmetric theory valid
at higher energy scales.  To give rise to spontaneous
supersymmetry breaking, the more fundamental theory must
contain other {\it sectors} that are responsible for this
effect.  Often the additional sectors include a
strongly-interacting super-Yang-Mills theory.
The most general effective
theory consistent with symmetry constraints typically
involves nonholomorphic quantities, determined by the
{\it K\"ahler potential.}  It is of interest to understand the
nonperturbative corrections that this potential may receive,
particularly if the corresponding dynamics are expected to play
an important role in determining qualitative features of the effective
theory; in addition to breaking supersymmetry,
nonperturbative effects are generally considered
to be important to {\it moduli stabilization}
(vacuum selection).

Thus, one hope for lattice supersymmetry
is that it would lead to simulations
that would provide further data on nonperturbative
aspects of supersymmetric
field theories, especially those that include
super-Yang-Mills.  However, the lattice
regulator generically breaks supersymmetry.
This is a natural consequence of the fact that
the supersymmetry algebra is embedded into the
super-Poincar\'e algebra, which involves both
rotations and translations.  Only a discrete subgroup
of the Poincar\'e group survives, so it is not
surprising that the regulator is not supersymmetric.
Thus the target theory is obtained by fine-tuning
the bare parameters of the lattice action---supplemented
by counterterms;
a considerable amount of work has been done
in this direction.  Early efforts in this direction can be
found in \cite{Curci:1986sm,Montvay:1994ze};
recent reviews with extensive references are given 
in \cite{Montvay:2001aj,Feo:2002yi}.  In the case
of pure 4d $\Ncal=1$ super-Yang-Mills, a gauge invariant
formulation with chiral lattice fermions
suffices to guarantee the correct continuum limit;
for example, simulations using domain wall fermions
have been performed \cite{Fleming:2000fa}.

It is worth noting, however,
that the discrete rotations and translations
that are preserved in a typical (isotropic,
hypercubic) latticization guarantee that in the
continuum limit the only relevant and marginal operators
that can appear are those that preserve the full
Poincar\'e invariance of the target theory.
It is of interest to explore whether or not
it is possible to realize something analogous
for the super-Poincar\'e group.
Some amount of {\it exact lattice supersymmetry---}i.e.,
a fermionic symmetry that relates lattice bosons and lattice
fermions and closes with the discrete subgroup of the Poincar\'e
group that is preserved by the lattice---would 
presumably allow for the target theory to be obtained in a
more controlled fashion.
Remarkably, in some recent examples possessing an
exact lattice supersymmetry, it has
been found that certain super-Yang-Mills theories
may be obtained from a lattice theory without
the need for fine-tuning; additionally,
other types of super-Yang-Mills theories may 
be obtained with less fine-tuning than would occur
in a naive discretization \cite{CKKU,Cohen:2003xe,Kaplan:2002wv}.
It is worth noting that other sorts 
of models with exact lattice supersymmetry have
been discussed in the literature by a few
groups:  supersymmetric
quantum mechanics \cite{Catterall:2000rv,Catterall:2001wx},
the 2d Wess-Zumino model \cite{Catterall:2001fr,Catterall:2001wx},
pure super-Yang-Mills using overlap fermions \cite{Maru:1997kh},
and direct constructions in the spirit
of the Ginsparg-Wilson relation\footnote{
L\"uscher first suggested that the {\it perfect action} approach
might be used to identify an analogue of
the Ginsparg-Wilson relation for lattice supersymmetry
\cite{Luscher:1998pq}.
This idea was worked out in noninteracting
examples in \cite{Aoyama:1998in,Bietenholz:1998qq}.
(An approach very similar
to \cite{Bietenholz:1998qq} has been applied
in \cite{So:1998ya}, yielding slightly different
expressions.)} have all been considered.
In some cases, the constructions have been related
to topological field theory \cite{Catterall:2003wd}.

\subsection{Deconstruction, orbifolds and lattice supersymmetry}
In this letter we will be interested in the
super-Yang-Mills constructions that lead to
a Euclidean lattice whose target
theory contains 8 supercharges \cite{CKKU}.
This is a generalization of the 4 supercharge
constructions of \cite{Cohen:2003xe}.
The method of building all such models is based on {\it deconstruction}
of extra dimensions \cite{Arkani-Hamed:2001ca,Hill:2000mu}.  The
corresponding interpretation in terms of the world-volume
theory of D-branes has led to the latticizations of
2d, 3d and 4d supersymmetric gauge theories.
While a complete latticization of spacetime is studied
here, partial latticizations also yield interesting
results; for example, the chiral anomalies
\cite{Giedt:2003xr} or instanton solutions \cite{Poppitz:2003uz} of 4d $N=2$
super-Yang-Mills can be equivalently described by
deconstructing one dimension to obtain a 3d $N=4$
product group ({\it quiver}) super-Yang-Mills theory that is an effective
latticization of one dimension.

The Euclidean spacetime
lattice constructions are all arrived at by {\it orbifold projections}
of super-Yang-Mills matrix models.\footnote{These {\it matrix
models} are obtained as 0d reductions
of 4d, 6d and 10d $\Ncal=1$ Euclidean super-Yang-Mills.}
I.e., in each case we quotient a
matrix model by some discrete symmetry group of the
theory.  Degrees of freedom that are not
invariant with respect to the combined action
of the orbifold generators are projected out.\footnote{For
a detailed discussion, we refer the reader to \cite{CKKU,Cohen:2003xe}.}
Following \cite{CKKU,Cohen:2003xe},
we will refer to the
``nonorbifolded'' matrix model as the {\it mother theory}
and the ``orbifolded'' matrix model as the {\it daughter theory.}
The (effective) lattice theory is obtained by studying
the daughter theory expanded about a nontrivial minimum
of its scalar potential; i.e., a point in its
{\it moduli space.}

\subsection{The fermion determinant}
In lattice theories containing fermions,
it is well-known that it is of great practical
importance that the fermion determinant (or more
generally, the Pfaffian), obtained
by integrating over the fermion degrees of freedom
in the partition function, be positive semi-definite.
For let $v$ represent the lattice bosons.  Then
having integrated out the lattice fermions, one obtains
an equivalent effective lattice action
($S_B$ is the bosonic part of the action before integrating
out fermions):
\beq
S_{\stxt{eff}}(v) &=& S_B(v) - \ln \det M(v).
\label{sefd}
\eeq
A positive semi-definite $\det M(v)$ yields a
real effective action, thus avoiding the inherent
problems of a complex action with respect to
estimating correlation functions by Monte
Carlo simulation.

However, in \cite{Giedt:2003ve} we found that the fermion
determinant was complex for the constructions of \cite{Cohen:2003xe},
which have (2,2) 2d $U(k)$ super-Yang-Mills as their
target theory.\footnote{In \cite{Giedt:2003ve}
we also showed that this was not in conflict
with the well-known positivity of the fermion determinant
in the mother theory \cite{Krauth:1998xh,Ambjorn:2000bf}.}  
Here we study whether or not a similar
problem exists in the constructions of \cite{CKKU}, which
have (4,4) 2d $U(k)$ super-Yang-Mills as their
target theory.

\subsection{Summary of results}
For the reader's convenience,
we now summarize the content of our work:
\bit
\item
In this letter we show that the lattice
theory with (4,4) 2d super-Yang-Mills
as its target \cite{CKKU}, obtained from
orbifolded supersymmetric matrix models, possesses a problematic
fermion determinant.
\item
Due to a ever-present fermion zeromode,
$\det M(\phi)\equiv 0$, i.e., for all boson configurations.
The zero eigenvalue can be factored
out in a controlled way in order to exhibit the
determinant for the other fermion modes.
In the daughter theory with (4,4) 2d $SU(2)$ super-Yang-Mills
as its target, we carry out this factorization (numerically).
\item
Once the zeromode fermion has been factored out,
{\it we find that the remaining product of eigenvalues
is generically nonzero with arbitrary complex phase.}
\item
We discuss the implications of
our results for lattice simulations of the
latticized (4,4) 2d super-Yang-Mills theories.
We emphasize that the complex action may be turned
into a virtue, in that it provides an interesting system
in which to study complex action simulation techniques---relevant
to lattice QCD at finite temperature and/or baryon density.
\item
We conclude by presenting some preliminary Monte Carlo results
for the distribution of the phase of the fermion determinant
when sampled in the {\it phase-quenched} distribution;
that is, when the determinant in \myref{sefd} is replaced
by its absolute value.  We discuss why this may be
relevant to the continuum limit.  Unfortunately,
for the small lattice that
we consider ($2 \times 2$),
we find that the phase distribution is essentially uniform
in it range $(-\pi,\pi]$.
\eit
In the remainder of this letter we will discuss various
details related to these results.

\section{Construction}
\subsection{Mother theory}
The action of the mother theory is that of a 6d $\to$ 0d
reduction of $U(kN^2)$ $\Ncal=1$ super-Yang-Mills:
\beq
S = -\frac{1}{4g^2} \tr \( [v_m,v_n][v_m,v_n] \)
+ \frac{1}{2g^2} \e_{ij} \tr \( \Psi^T_i C \bbar{\Sigma}_m
[v_m,\Psi_j] \)
\label{mtha}
\eeq
where $\e = i \s_2$, $v_m = v_m^\alpha T^\alpha$, $\Psi_i =
\Psi_i^\alpha T^\alpha \; (i=1,2)$ with $T^\alpha = ({\bf 1},T^a)$ a
Hermitian basis for the generators of $U(kN^2)$.
Each of the two $\Psi_i$ is a 4-component fermion.
The $4 \times 4$ matrices $\bbar{\Sigma}_m$
are components in the construction
of the 6d (Euclidean) Clifford algebra, and $C$ is a
charge conjugation matrix.  For further details
we refer the reader to \cite{CKKU}.

\subsection{Orbifold to daughter theory}
In addition to the $U(kN^2)$ gauge invariance, the mother
theory possesses an $SO(6)_E$ Euclidean invariance group
and an $SU(2)_R$ chiral R-symmetry group.  From the
invariance group $SO(6)_E \otimes SU(2)_R$
one isolates a $U(1)_1 \otimes U(1)_2$ subgroup,
with corresponding generators $r_1$ and $r_2$.  The
bosons and the fermions of the mother theory are written
in a basis with well-defined $r_1, r_2$ charges.  In
the construction of \cite{CKKU} this basis is:\footnote{Our
lattice boson notation is related to 
that of \cite{CKKU} by $(x,y,z) \equiv
(z_1,z_2,z_3)$.}
\beq
\Psi_1^T = (\lambda, \xi_1, \xi_2, \xi_3), \qquad
\Psi_2^T = (\chi, \psi_1, \psi_2, \psi_3),
\label{mtfm}
\eeq
\beq
C \bbar{\Sigma}_m v_m = \begin{pmatrix}
0 & \xb & \yb & \zb \cr
- \xb & 0 & z & -y \cr
- \yb & -z & 0 & x \cr
- \zb & y & -x & 0 \cr
\end{pmatrix}
\eeq
The $r_1, r_2$ charges are given in Table 1 of \cite{CKKU};
we will not need them here.  All that is important is
that the bosons and fermions (denoted collectively by $\Phi$) of the
mother theory are subjected to a projection
with respect to a $Z_N \otimes Z_N$ subgroup
of $U(1)_1 \otimes U(1)_2$, together with a nontrivial embeddeding into
the gauge group $U(kN^2)$.  That is,
we keep only fields that satisfy
\beq
\Phi \equiv e^{2\pi i r_a/N} {\cal C}_a \Phi {\cal C}_a^{-1}
\label{proo}
\eeq
where ${\cal C}_a$ are generators of
a $Z_N \otimes Z_N$ subgroup of $U(kN^2)$.
This breaks the gauge group down to $U(k)^{N^2}
= \bigotimes_{m_1,m_2=1}^N U(k)_{m_1,m_2}$,
corresponding the $U(k)$ gauge invariance of an $N \times N$
lattice theory.  Bosons and fermions associated
with a single factor $U(k)_{m_1,m_2}$ of $U(k)^{N^2}$
are interpreted as site variables; in an appropriate
basis, the other surviving
bosons and fermions are charged with respect to 2 factors
and are therefore interpreted as link variables.

\section{Fermion action}
Here we examine the fermion determinant for the
daughter theory.  Expanded about the chosen point
in moduli space, it is the fermion determinant
for the lattice theory.  We find it convenient to define
\beq
\tr (T^\mu T^\nu T^\rho) = \tilde N t^{\mu \nu \rho}, \qquad
t_{\mbf, \nbf}^{\mu \nu \rho} = \delta_{\mbf,\nbf} t^{\mu \nu \rho}
\eeq
where $\tilde N$ is an overall normalization that may
be chosen as seems convenient, and
$\mbf=(m_1,m_2)$ labels sites on a 2d
square lattice; $\ibf=(1,0)$ and $\jbf=(0,1)$ are unit vectors in the
two directions.

The fermion matrix depends on bosons
\beq
x_\mbf = x_\mbf^\mu T^\mu, \qquad
y_\mbf = y_\mbf^\mu T^\mu, \qquad
z_\mbf = z_\mbf^\mu T^\mu
\eeq
as well as conjugates $\xb_\mbf^\mu = (x_\mbf^\mu)^\dagger$,
etc.  The fermion action can be written in the form
\beq
S_F = - \frac{\tilde N\sqrt{2}}{g^2}
\( \psi_{1,\mbf}^\mu ~,~ \psi_{2,\mbf}^\mu ~,~
\psi_{3,\mbf}^\mu ~,~ \chi_\mbf^\mu \)
\cdot M_{\mbf \nbf}^{\mu \rho} \cdot
\begin{pmatrix} \xi_{1, \nbf}^\rho \cr
\xi_{2, \nbf}^\rho \cr \xi_{3, \nbf}^\rho
\cr \lambda_\nbf^\rho \cr \end{pmatrix}
\label{haht}
\eeq
The elements of the fermion matrix that follow from the
expressions of \cite{CKKU} and the conventions chosen here
are:
\beq
(M_{\mbf \nbf}^{\mu \rho})_{1,1} &=&
(M_{\mbf \nbf}^{\mu \rho})_{2,2} \; = \;
(M_{\mbf \nbf}^{\mu \rho})_{3,3} \; = \;
(M_{\mbf \nbf}^{\mu \rho})_{4,4} \; = \; 0,
\nnn
(M_{\mbf \nbf}^{\mu \rho})_{1,2} &=&
- \tmr_{\mbf,\nbf} z_{\nbf+\ibf}^\nu + \tmn_{\mbf,\nbf} z_\nbf^\nu ,
\qquad
(M_{\mbf \nbf}^{\mu \rho})_{1,3} =
\tmr_{\mbf,\nbf} y_{\nbf+\ibf}^\nu - \tmn_{\mbf,\nbf+\jbf} y_\nbf^\nu ,
\nnn
(M_{\mbf \nbf}^{\mu \rho})_{1,4} &=&
\tmr_{\mbf,\nbf} \xb_{\nbf}^\nu - \tmn_{\mbf,\nbf-\ibf} \xb_\mbf^\nu ,
\qquad
(M_{\mbf \nbf}^{\mu \rho})_{2,1} =
\tmr_{\mbf,\nbf} z_{\nbf+\jbf}^\nu - \tmn_{\mbf,\nbf} z_\nbf^\nu ,
\nnn
(M_{\mbf \nbf}^{\mu \rho})_{2,3} &=&
- \tmr_{\mbf,\nbf} x_{\nbf+\jbf}^\nu + \tmn_{\mbf,\nbf+\ibf} x_\nbf^\nu ,
\qquad
(M_{\mbf \nbf}^{\mu \rho})_{2,4} =
\tmr_{\mbf,\nbf} \yb_{\nbf}^\nu - \tmn_{\mbf,\nbf-\jbf} \yb_\mbf^\nu ,
\nnn
(M_{\mbf \nbf}^{\mu \rho})_{3,1} &=&
- \tmr_{\mbf,\nbf} y_{\nbf}^\nu + \tmn_{\mbf,\nbf+\jbf} y_\nbf^\nu ,
\qquad
(M_{\mbf \nbf}^{\mu \rho})_{3,2} =
\tmr_{\mbf,\nbf} x_{\nbf}^\nu - \tmn_{\mbf,\nbf+\ibf} x_\nbf^\nu ,
\nnn
(M_{\mbf \nbf}^{\mu \rho})_{3,4} &=&
\tmr_{\mbf,\nbf} \zb_{\nbf}^\nu - \tmn_{\mbf,\nbf} \zb_\nbf^\nu ,
\qquad
(M_{\mbf \nbf}^{\mu \rho})_{4,1} =
- \tmr_{\mbf,\nbf} \xb_{\nbf+\jbf}^\nu
+ \tmn_{\mbf,\nbf-\ibf} \xb_\mbf^\nu ,
\nnn
(M_{\mbf \nbf}^{\mu \rho})_{4,2} &=&
- \tmr_{\mbf,\nbf} \yb_{\nbf+\ibf}^\nu
+ \tmn_{\mbf,\nbf-\jbf} \yb_\mbf^\nu ,
\qquad
(M_{\mbf \nbf}^{\mu \rho})_{4,3} =
- \tmr_{\mbf,\nbf} \zb_{\nbf+\ibf+\jbf}^\nu
+ \tmn_{\mbf,\nbf} \zb_\nbf^\nu .
\eeq

\subsection{Fermion zeromode}
The mother theory fermion modes $\lambda = \lambda^\alpha T^\alpha$
that appear in \myref{mtfm} are $U(1)_1 \otimes U(1)_2$
neutral; that is, they have $r_1=r_2=0$ in \myref{proo}.
As a consequence, the surviving parts correspond to $T^\alpha$
that are nothing but the Cartan subalgebra of the mother theory,
including the operator $T^0 = {\bf 1}$.  Because of the commutator
in \myref{mtha}, this fermion mode $\lambda^0$ disappears
from the action.  Thus it corresponds to an ever-present
zeromode in both the mother theory and the daughter theory.

This zeromode eigenvalue of the daughter theory
can be factored out following the method used
in \cite{Giedt:2003ve}.  We
deform the fermion matrix appearing in \myref{haht} according to
\beq
M \to M_\e \equiv M + \e {\bf 1}_{N_f}
\eeq
where $N_f = 4 k^2 N^2$ is the dimensionality of the fermion matrix
and $\e \ll 1$ is a deformation parameter that we will
eventually take to zero.  We factor out the zero mode
through the definition
\beq
&& \hat M(0) = \lim_{\e \to 0^+} \hat M(\e), \quad
\hat M(\e) \equiv \e^{-1/N_f} M_\e  \; \Rightarrow \;
\det \hat M(0) = \lim_{\e \to 0^+} \e^{-1} \det M_\e ~ .
\quad
\eeq

If this deformation is added to the action, it explicitly
breaks the exact lattice supersymmetry and gauge invariance.
This infrared regulator could be removed in the continuum
limit, say, by taking $\epsilon a \ll N^{-1}$.  Noting that
$L = Na$ is the physical size of the lattice, the
equivalent requirement is that $\epsilon \ll L^{-1}$
be maintained as $a \to 0$, for fixed $L$.  Thus in the
thermodynamic limit ($L \to \infty$), the deformation is
removed.  The parameter $\epsilon$ is a soft
infrared regulating mass, and is quite analogous to the
soft mass $\mu$ introduced by Cohen et al.~\cite{CKKU}
in their Eq.~(1.2) to control the bosonic zeromode of
the theory.\footnote{See also \myref{sboa} below.}  
In the same way that the deformation introduced with $\mu$
does not modify the final results of the renormalization
analysis of Section 3.4 of \cite{CKKU}, our $\epsilon$
does not modify the result of the quantum continuum limit.
The essence of the argument is that we have introduced a
vertex that will be proportional to the dimensionless
quantity $g_2^2 \epsilon a^3 \ll g_2^2 a^3 / L$, where
$g_2$ is the 2d coupling constant.  Such
contributions to the operator coefficients $C, \hat C$
in Eq.~(3.29) of \cite{CKKU} vanish in the thermodynamic
limit. Because the target theory is super-renormalizable,
we are assured that the perturbative power counting
arguments are reliable and the correct continuum limit is
obtained.

We have studied the convergence of
$\det \hat M(\e) \to \det \hat M(0)$.  Indeed,
we find that the convergence
is rapid and that a reliable estimate for $\det \hat M(0)$
can be obtained in this way.  As a check, we have computed
the eigenvalues of the undeformed matrix $M$,
using the math package Maple, for
a subset of 10 of the random boson configurations studied below.
We find that the product of nonzero eigenvalues
agrees with $\det \hat M(0)$ in magnitude and phase
to within at least 5 significant digits in each of the 10 cases.

\subsection{$U(2)$ fermion determinant}
In our numerical work,
we specialize to $U(2)$.  In notation introduced
above, we choose
\beq
&& T^\mu = ( \obf_2, \s^a ), \qquad
\tr (T^\mu T^\nu T^\rho) = 2 t^{\mu \nu \rho} \qquad
\Rightarrow \nnn
&& t^{000}=1, \qquad t^{\underline{a00}} = 0, \qquad
t^{\underline{ab0}} = \delta^{ab}, \qquad t^{abc} = i\e^{abc}
\eeq
where underlining of indices indicates that all permutations are
to be taken.

The bosons have components
\beq
x_\mbf = x_\mbf^0 \obf_2 + x_\mbf^a \s^a, \qquad
y_\mbf = y_\mbf^0 \obf_2 + y_\mbf^a \s^a, \qquad
z_\mbf = z_\mbf^0 \obf_2 + z_\mbf^a \s^a .
\label{qqre}
\eeq
The lattice theory with lattice spacing $a$
is obtained by expansion about
a particular point in moduli space:
\beq
x_\mbf^0 = \frac{1}{a\sqtw} + \cdots, \qquad
y_\mbf^0 = \frac{1}{a\sqtw} + \cdots, \qquad
\label{xyv}
\eeq
where $\cdots$ represent the quantum fluctuations
and all other bosons are expanded about the orgin.
For this reason, in our study of $\det \hat M(0)$
we scan over a Gaussian distribution
where $x_\mbf^0,y_\mbf^0$ have a nonzero mean $1/a\sqtw \equiv 1$.
The remainder of the bosons are drawn with mean
zero.  All bosons are taken from distributions
with unit variance.

\begin{figure}
\begin{center}
\includegraphics[height=5.0in,width=2.0in,angle=90]{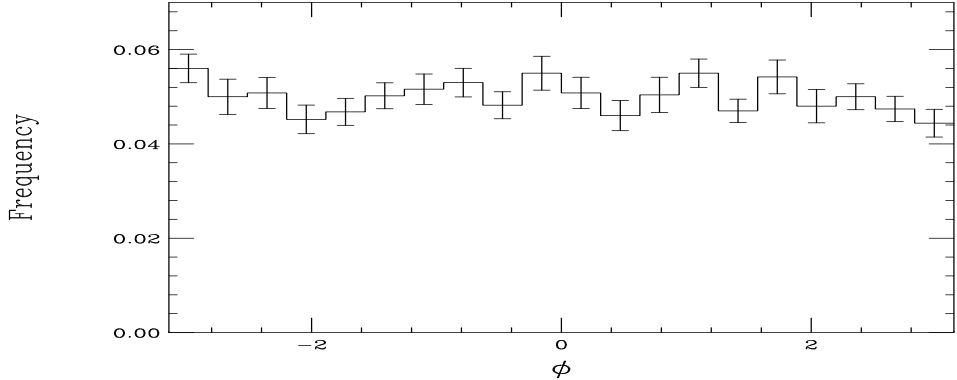}
\end{center}
\caption{Average frequency distribution $F(\phi)$
for $\phi = \arg \det \hat M(0)$,
for 5000 random (Gaussian) draws, binned into intervals of $\pi/10$.
Data was arranged into 50 blocks of 100 draws to estimate errors.
The distribution of $\phi$ is seen to be, within errors, uniform.
These results are for
the $U(2)$ lattice theory, with $2 \times 2$ lattice.
}
\label{f1}
\end{figure}

For a set of $5000$ draws on the bosons of a
$2 \times 2$ lattice, we have extrapolated to $\e \to 0$ and
binned $\phi \equiv \arg \det \hat M(0)$ over its range, with
20 bins of size $\pi/10$.  In Fig.~\ref{f1}
we show the frequency for each bin, as a fraction of
the total number of draws.
It can be seen that once the zeromode eigenvalue is factored out,
the product of the nonzero eigenvalues has arbitrary phase
and that within statistical errors
the distribution $F(\phi)$ is uniform:  $F(\phi) \approx 1/20$.
Consequently, the effective lattice action [cf.~\myref{sefd}]
is complex, with the phase of the fermion determinant
a field-dependent quantity.
This is to be contrasted with the observed reality of the
fermion determinant in the mother theory \cite{Krauth:1998xh}.
Presumably the orbifold projection does not commute
with the conjugation operator that guarantees this
reality in the mother theory.  Indeed, it is not difficult
to check that the orbifold projection operator does
not commute with the usual conjugation operator that
is involved in establishing the hermiticity of the $\Ncal=1$
6d Minkowski spacetime action.  Thus it is not surprising
that the fermion determinant in the daughter theory is
not real.

\section{Re-weighting}
It is worthwhile to explore whether or
not the complex phase can be overcome for the purposes
of simulation.  A typical approach would be to compute
averages of an operator ${\cal O}$ from the {\it re-weighting}
identity:
\beq
\vev{ {\cal O} } = \frac{\bigvev{ {\cal O} e^{i \phi} }_{p.q.}}
{ \bigvev{ e^{i \phi} }_{p.q.}}
\label{opq}
\eeq
Here, $\phi= \arg \det \hat M(0)$, as above,
and ``p.q.'' indicates {\it phase-quenching:}
expectation values are computed with the replacement $\det \hat M(0)
\to |\det \hat M(0)|$.  Thus the effective bosonic action
\beq
S_{\stxt{p.q.}} = S_B - \ln |\det \hat M(0)|
\label{spq}
\eeq
is used to generate the {\it phase-quenched ensemble}
by standard Monte Carlo techniques.
However, it is well-known that this tends to suffer efficiency
problems:  the number of configurations required to get
an accurate estimate for, say,
$\bigvev{ \exp ( i \phi ) }_{p.q.}$,
grows like $\exp (\Delta F \cdot N_f^2)$.  Here,
$\Delta F$ is the difference in free energy densities between the
full ensemble and the phase-quenched ensemble.
Recall that $N_f$ is the dimensionality of the fermion
matrix.  It has been recently suggested how
difficulties with this re-weighting
approach might be surmounted by {\it distribution
factorization} techniques \cite{AN01a}.
Indeed, this method has been fruitfully applied
in some other systems with complex action,
some of which are quite similar to that studied
here \cite{Ambjorn:2002pz,Azcoiti:2002vk,Nis03a}.
It would be interesting to
see what progress may be made in the present context
by applying this method.  We are currently examining
this idea and hope to report on it in a later publication.

However, it is also possible that in the
continuum limit the phase-quenched distribution
is sharply peaked at a value $\phi_0$ of $\phi$.
It follows that
\beq
\vev{ {\cal O} } \approx \frac{e^{i \phi_0} \bigvev{ {\cal O}  }_{p.q.}}
{e^{i \phi_0}  \bigvev{ 1 }_{p.q.}}
= \bigvev{ {\cal O}  }_{p.q.}
\eeq
Thus it may be that the phase-quenched ensemble gives
a good estimation in the continuum limit.  For this
reason it is of interest to study the distribution
of $\phi$ as determined by the phase-quenched ensemble
as a function of the lattice spacing.  We next report
results of a preliminary study of this distribution
using Monte Carlo techniques.

\section{Phase-quenched distribution}
We generate a set of configurations, updating using
the Metropolis algorithm applied to \myref{spq}.
In this way, we sample the phase-quenched ensemble and
estimate the corresponding distribution of $F_{p.q.}(\phi)$.
For this purpose we need the
bosonic action of the daughter theory.  We specialize again to the
$U(2)$ case and introduce the notation $\tr ( T^\mu T^\nu T^\rho T^\lambda )
= 2 t^{\mu \nu \rho \lambda}$.  Then refering to \cite{CKKU},
the bosonic action is given by
\beq
S_B &=& \frac{2}{g^2} t^{\mu \nu \rho \lambda} \sum_{\nbf} \left[
\half (\xb_{\nbf-\ibf}^\mu x_{\nbf-\ibf}^\nu - x_{\nbf}^\mu \xb_{\nbf}^\nu
+ \yb_{\nbf-\jbf}^\mu y_{\nbf-\jbf}^\nu - y_\nbf^\mu \yb_\nbf^\nu
+ \zb_\nbf^\mu z_\nbf^\nu - z_\nbf^\mu \zb_\nbf^\nu ) \right. \nnn
&& \times ( \xb_{\nbf-\ibf}^\rho x_{\nbf-\ibf}^\lambda
- x_{\nbf}^\rho \xb_{\nbf}^\lambda
+ \yb_{\nbf-\jbf}^\rho y_{\nbf-\jbf}^\lambda
- y_\nbf^\rho \yb_\nbf^\lambda
+ \zb_\nbf^\rho z_\nbf^\lambda - z_\nbf^\rho \zb_\nbf^\lambda) \nnn
&& + 2(x_\nbf^\mu y_{\nbf+\ibf}^\nu - y_\nbf^\mu x_{\nbf+\jbf}^\nu)
(\yb_{\nbf+\ibf}^\rho \xb_\nbf^\lambda - \xb_{\nbf+\jbf}^\rho \yb_\nbf^\lambda)
+ 2(y_\nbf^\mu z_{\nbf+\jbf}^\nu - z_\nbf^\mu y_\nbf^\nu)
(\zb_{\nbf+\jbf}^\rho \yb_\nbf^\lambda - \yb_\nbf^\rho \zb_\nbf^\lambda)
\nnn && \left. + 2(z_\nbf^\mu x_\nbf^\nu - x_\nbf^\mu z_{\nbf+\ibf}^\nu)
(\xb_\nbf^\rho \zb_\nbf^\lambda - \zb_{\nbf+\ibf}^\rho \xb_\nbf^\lambda)
+ \frac{a^2 \mu^2}{2} (x_\nbf^\mu \xb_\nbf^\nu x_\nbf^\rho \xb_\nbf^\lambda
+ y_\nbf^\mu \yb_\nbf^\nu y_\nbf^\rho \yb_\nbf^\lambda)
\right] \nnn &&
+ \frac{\mu^2}{g^2} \sum_\nbf \[ 2 z_\nbf^\mu \zb_\nbf^\mu
- x_\nbf^\mu \xb_\nbf^\mu - x_\nbf^\mu \xb_\nbf^\mu \] + {\rm const.}
\label{sboa}
\eeq
with an implied sum over repeated superscripts.
The parameter $\mu$ is a soft supersymmetry breaking
mass that is inserted to stabilize the 
vacuum expectation values \myref{xyv}
and is tuned to zero as the infinite volume limit is taken.
In particular, in our simulations we are careful to respect the relation
\beq
ga^2 \ll \mu a N \lappeq 1 .
\label{uirt}
\eeq
Further details on this deformation of the
bosonic action may be found in \cite{CKKU}.

We have studied the distribution $F_{p.q.}(\phi)$
on small lattices for a few
choices of the bare parameters.  It is convenient
to rewrite these parameters in terms of physically
meaningful quantities with dimensions of length:
\beq
a, \qquad g_2^{-1} = (ga)^{-1}, \qquad L = Na, \qquad \mu^{-1}.
\eeq
We expect that $g_2^{-1}$ is a rough measure
of the correlation length for the system.  Thus, we
anticipate that systematics due to latticization and 
working at finite volume are kept to a minimum provided
\beq
a \ll g_2^{-1} \ll L .
\label{uire}
\eeq
The effects of supersymmetry breaking
in the infrared are expected to be negligible provided
$g_2^{-1} \ll \mu^{-1}$.  In the our simulations, we have
examined the following parameter sets:
\beq
(N;a,g_2^{-1},L,\mu^{-1}) &=&  (2; 1/8, 8, 1/4, 1/2), 
\quad (2; 1/20, 40, 1/10, 4/3), \nnn &&  (3; 1/20, 40, 3/20, 2), 
\quad (3; 2/3, 1, 2, 2), \nnn && (4; 1/2, 1, 2, 2).
\eeq
The first three sets suffer from extreme finite volume
effects, but the effects of discretization are 
expected to be rather small since $a g_2 \leq 1/64$.
Supersymmetry breaking is expected to be
small in the second and third sets since $\mu / g_2 \leq 3/160$.
The last two cases respect a weaker version of
the ``physical'' constraint \myref{uire}, namely
$a < g_2^{-1} < L$, and simultaneously a weaker version
of \myref{uirt}, namely $ga^2 < \mu a N = 1$.
(The conditions were of necessity weakened, due to the small
lattices we are presently working with.)
In all cases, we find that the distribution is uniform;
that is, up to statistical fluctuations, the results
are indistinguishable from those of Fig.~\ref{f1}.

\section{Conclusions}
Naturally we would like to study the phase-quenched
distribution $F_{p.q.}(\phi)$ on larger lattices
where we can better approach the continuum limit.
However, these simulations are rather demanding
since they involve a fermion determinant; we
leave such an effort to future work, as it will
take some time to accumulate the necessary lattice
data and develop efficient algorithms.  Moreover, it
is not at all clear that the phase of the determinant over {\it all}
lattice fermion modes, over {\it all} nonzero measure boson
configurations in the phase-quenched ensemble,
is entirely relevant to the phase of the
continuum limit fermion determinant.  Only those modes that have
a significant overlap with the low energy interpolating
fields are expected to have physical meaning.
The other modes integrate out without significantly
affecting correlation functions of physical operators.
This is closely related to the distribution
factorization method mentioned above, in the
case where the operator ${\cal O}$ appearing
in \myref{opq} corresponds to a good approximation
to an operator in the target theory with
external momentum scales well below $a^{-1}$.
Research in this direction is underway.

Alternatively, it may be hoped that a different orbifold
may preserve the reflection positivity of the mother theory
while still producing an effective lattice theory with
exact supersymmetry.  We suspect that
the fermion determinant would be positive
in such a construction.  In the present constuction
it seems that reflection positivity is only
recovered in the continuum limit.  We are
also pursuing this notion in ongoing research.

The complex action problem encountered here provides
an amusing opportunity.  Some exact results are
available for the (4,4) 2d super-Yang-Mills theories.
In particular, the 1-loop effective action has
been determined \cite{Smilga:2002hu},
modulo higher derivative terms.  It would be 
interesting to compare the results of
complex action simulation techniques 
to analytic results such as these.\footnote{Exacting tests of
the simulation techniques would of course require reliable
{\it nonperturbative} analytic results in the 
(4,4) 2d super-Yang-Mills theories.  This should be
possible by a dimensional reduction of the Seiberg-Witten
results \cite{Seiberg:1994rs}.}
It may be that the supersymmetric lattice actions proposed here 
provide an independent
check on Monte Carlo methods that have been suggested in the
simulation of QCD at finite temperature and baryon
density.

\vspace{15pt}

\noindent
{\bf \Large Acknowledgements}

\vspace{5pt}

\noindent
The author would like to than David B. Kaplan for a preliminary
version of \cite{CKKU} and other helpful communications.  Appreciation
is also owed to Erich Poppitz for comments.
This work was supported
by the National Science and Engineering Research Council of Canada.

\end{document}